\documentclass[aps,prx,twocolumn,showpacs,psfig,superscriptaddress,longbibliography]{revtex4-1}

\usepackage{times}
\usepackage{graphicx}
\usepackage{float}
\usepackage{latexsym,amsmath,amssymb,bm,euscript}
\usepackage{color}
\usepackage{subfigure}
\usepackage{epstopdf}
\usepackage[colorlinks=true,linkcolor=blue,citecolor=blue]{hyperref}
\usepackage{soul}
\usepackage[normalem]{ulem}
\usepackage{mathrsfs}
\usepackage{amsmath}
\usepackage{lettrine}
\usepackage{bbding}
\usepackage{xspace}
\usepackage{textcomp}
\usepackage{textcase}
\usepackage{setspace}
\usepackage{multirow}
\usepackage[version=4]{mhchem}

\def\para{\ensuremath{/\kern -0.8em /}\xspace}

\def\beqn{\begin{eqnarray}}
\def\eeqn{\end{eqnarray}}
\def\beq{\begin{equation}}
\def\eeq{\end{equation}}

\newcommand{\Beq}{\begin{eqnarray*} }
\newcommand{\Eeq}{\end{eqnarray*} }
\newcommand{\Bmat}{\left(\begin{matrix}}
\newcommand{\Emat}{\end{matrix}\right)}

\begin{document}

\title{Gapless spinon excitations emerging from a multipolar transverse field in the triangular-lattice Ising antiferromagnet NaTmSe\textsubscript{2}}

\author{Zheng Zhang}
\affiliation{Beijing National Laboratory for Condensed Matter Physics, Institute of Physics, Chinese Academy of Sciences, Beijing 100190, China}

\author{Jinlong Jiao}
\affiliation{Department of Physics and Astronomy, Shanghai Jiao Tong University, Shanghai 200240, China}

\author{Weizhen Zhuo}
\affiliation{Beijing National Laboratory for Condensed Matter Physics, Institute of Physics, Chinese Academy of Sciences, Beijing 100190, China}
\affiliation{School of Physical Science and Technology, Lanzhou University, Lanzhou 730000, China}

\author{Mingtai Xie}
\affiliation{Beijing National Laboratory for Condensed Matter Physics, Institute of Physics, Chinese Academy of Sciences, Beijing 100190, China}
\affiliation{School of Physical Science and Technology, Lanzhou University, Lanzhou 730000, China}

\author{D.T. Adroja}
\affiliation{ISIS Neutron and Muon Facility, STFC Rutherford Appleton Laboratory, Chilton, Didcot Oxon, OX11 0QX, United Kingdom}
\affiliation{Highly Correlated Matter Research Group, Physics Department, University of Johannesburg, PO Box 524, Auckland Park 2006, South Africa}

\author{Toni Shiroka}
\affiliation{Laboratory for Muon-Spin Spectroscopy, Paul Scherrer Institut, Villigen PSI, Switzerland}
\affiliation{Laboratorium für Festkörperphysik, ETH Zürich, CH-8093 Zürich, Switzerland}

\author{Guochu Deng}
\affiliation{Australian Centre for Neutron Scattering, Australian Nuclear Science and Technology Organization, New Illawarra Road, Lucas Heights, NSW 2234, Australia}

\author{Anmin Zhang}
\affiliation{School of Physical Science and Technology, Lanzhou University, Lanzhou 730000, China}

\author{Feng Jin}
\affiliation{Beijing National Laboratory for Condensed Matter Physics, Institute of Physics, Chinese Academy of Sciences, Beijing 100190, China}

\author{Jianting Ji}
\affiliation{Beijing National Laboratory for Condensed Matter Physics, Institute of Physics, Chinese Academy of Sciences, Beijing 100190, China}

\author{Jie Ma}
\email{jma3@sjtu.edu.cn}
\affiliation{Department of Physics and Astronomy, Shanghai Jiao Tong University, Shanghai 200240, China}

\author{Qingming Zhang}
\email{qmzhang@iphy.ac.cn}
\affiliation{Beijing National Laboratory for Condensed Matter Physics, Institute of Physics, Chinese Academy of Sciences, Beijing 100190, China}
\affiliation{School of Physical Science and Technology, Lanzhou University, Lanzhou 730000, China}

\begin{abstract}
	The triangular-lattice quantum Ising antiferromagnet is a promising platform for realizing Anderson's quantum spin liquid, though finding suitable materials to realize it remains a challenge. 
	Here, we present a comprehensive study of \ce{NaTmSe2} using magnetization, specific heat, neutron scattering, and muon spin relaxation, combined with theoretical calculations. 
	We demonstrate that \ce{NaTmSe2} realizes the transverse field Ising model and quantitatively determine its exchange parameters.
	Our results reveal a multipolar spin-polarized state coexisting with a dipolar spin-disordered state. These states feature gapless spinon excitations mediated by the multipolar moments.
	The study shows how multiple types of magnetism can emerge in distinct magnetic channels (dipolar and multipolar) within a single magnet, advancing our understanding of spin-frustrated Ising physics and opening pathways for novel quantum computing applications.
\end{abstract}
\date{\today}
\maketitle

\textit{Introduction}-----
The Ising model is central to understanding phase transitions and exotic spin excitations \cite{RevModPhys.36.856,RevModPhys.39.883}. 
In the transverse field Ising model (TFIM), the noncommutativity $\left[ S_{x}, S_{z} \right] \neq 0$ precludes a classical description and requires a fully quantum treatment.
The transverse field drives quantum phase transitions from ordered to disordered states. 
For example, in the one-dimensional Ising chain \ce{SrCo2V2O8}, the Néel order is suppressed by a transverse field, 
generating a quantum critical point \cite{PhysRevLett.123.067203}. 
Remarkably, $E_{8}$ particles emerge in \ce{BaCo2V2O8} under transverse fields \cite{doi:10.1126/science.1180085, PhysRevB.96.024439, PhysRevLett.123.067202, PhysRevLett.127.077201} due to the realization of an integrable quantum field theory with $E_{8}$ symmetry at the critical point of a quantum phase transition.
On a square lattice, TFIMs with next-nearest-neighbor (NNN) spin interactions also exhibit field-induced quantum phase transitions due to quantum fluctuations \cite{Oitmaa_2020, WANG2021168522, PhysRevB.94.214419}.

Spin frustration enhances quantum fluctuations, leading to exotic quantum effects and excitations. 
Examples include the kagome compound \ce{ZnCu3(OH)6Cl2} \cite{RN106}, the triangular lattice \ce{YbMgGaO4} \cite{li2015gapless,li2015rare}, and recent rare-earth chalcogenides \cite{Liu_2018,ChinPhysLett.41.117505,RN103,zhuo_magnetism_2024,PhysRevX.11.021044}. 
Exchange frustration in the Kitaev model on a honeycomb lattice arises from anisotropic, bond-dependent spin interactions \cite{Kitaev2006}. 
Materials such as $\alpha$-\ce{RuCl3} \cite{PhysRevLett.119.227202,PhysRevLett.119.227208}, \ce{Na2Co2TeO6} \cite{RN23}, \ce{YbCl3} \cite{Hao2020,sala_field-tuned_2023,matsumoto_quantum_2024}, 
and \ce{YbOCl} \cite{PhysRevResearch.4.033006,PhysRevResearch.6.033274,PhysRevResearch.6.033274} have attracted significant attention.

The combination of geometric spin frustration and TFIM provides a promising platform for exploring exotic spin phases and excitations of Anderson's quantum spin liquid (QSL) on a triangular lattice \cite{ANDERSON1973153,doi:10.1080/14786439808206568}.
However, realizing the spin-frustrated TFIM is challenging, as it requires extremely high spin anisotropy to suppress the in-plane spin components.
Rare-earth ions, with their strong spin-orbit coupling (SOC), are promising candidates for realizing TFIM systems  \cite{PhysRevB.98.045119,PhysRevResearch.2.043013}.

\ce{TmMgGaO4} \cite{CEVALLOS2018154,PhysRevB.103.064424,PhysRevX.10.011007,PhysRevResearch.2.043013,RN109,RN116,RN122,QIN202238} is a well-studied rare-earth compound featuring the TFIM, where the Ising term is described by the dipolar operator $S^{z}$, while the transverse field couples to a higher-order multipolar operator.
Although multipolar moments remain undetectable in conventional measurements \cite{PhysRevResearch.2.043013}, their interplay with dipolar moments leads to a clock-ordered ground state \cite{RN109}.
\ce{NaTa7O19} is another triangular-lattice antiferromagnet that exhibits Ising-like correlations but no magnetic ordering, as demonstrated by magnetization, electron spin resonance, and muon spin relaxation ($\mu$SR) experiments \cite{arh_ising_2022}. 
Similarly, \ce{PrMgAl11O19} \cite{PhysRevB.109.165143,PhysRevB.110.134401} and \ce{PrZnAl11O19} \cite{PhysRevB.106.134428} exhibit strong Ising magnetism
and disordered ground states with gapless excitations.

\begin{figure}[t!]
	\includegraphics[angle=0,width=1\linewidth]{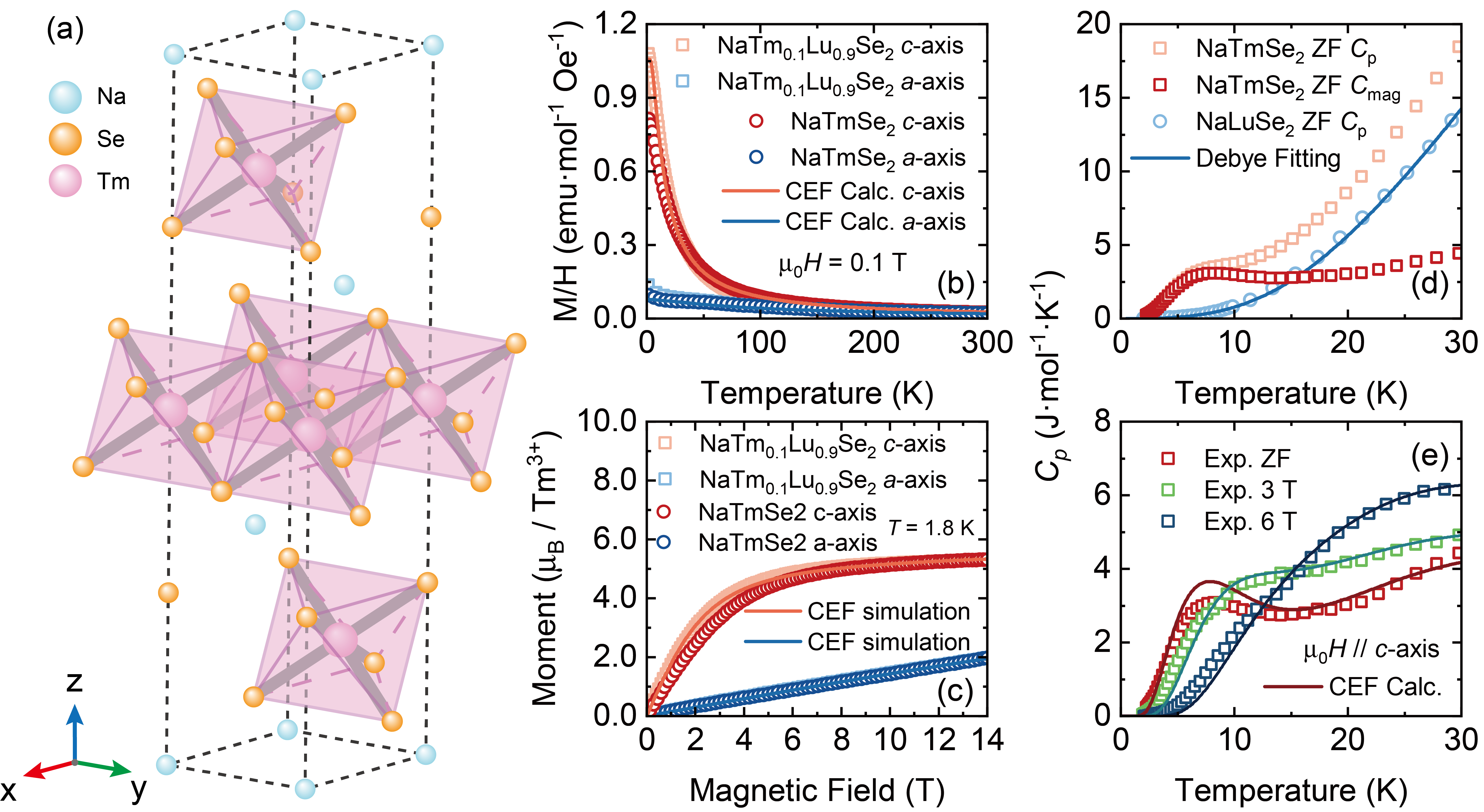}
	\renewcommand{\figurename}{\textbf{Fig. }}
	\caption{
		\textbf{Crystal structure and thermodynamic properties of \ce{NaTmSe2}.} 
		\textbf{(a)} Crystal structure showing \ce{TmSe6} octahedra forming two-dimensional triangular layers of magnetic \ce{Tm^{3+}} ions in the $ab$-plane, separated by \ce{Na^{+}} layers.
		\textbf{(b)} Temperature dependent magnetization measured under a magnetic field of 0.1 T along the $c$-axis and $a$-axis for \ce{NaTmSe2} (circles) and \ce{NaTm_{0.1}Lu_{0.9}Se2} (squares).
		\textbf{(c)} Magnetic field dependent magnetization at $T$ $=$ 1.8 K up to 14 T along the $c$-axis and $a$-axis for \ce{NaTmSe2} (circles) and \ce{NaTm_{0.1}Lu_{0.9}Se2} (squares).
		Solid lines in \textbf{(b)} and \textbf{(c)} represent CEF calculations.
		\textbf{(d)} Specific heat of non-magnetic \ce{NaLuSe2} (light blue circles) with Debye fitting (blue line). Raw and magnetic specific heat of \ce{NaTmSe2} are shown as orange and red squares, respectively.
		\textbf{(e)} Magnetic specific heat of \ce{NaTmSe2} under different magnetic field along $c$-axis (squares) with corresponding CEF calculations (solid lines).
		}
	\label{Fig:Intro}
\end{figure}

We introduce \ce{NaTmSe2}, a member of rare-earth chalcogenides. 
The \ce{Tm^{3+}} ions in both \ce{TmMgGaO4} and \ce{NaTmSe2} share the $D_{3d}$ point group symmetry and a similar crystalline electric field (CEF) environment, indicating the presence of Ising magnetism in \ce{NaTmSe2}. 
\ce{NaTmSe2} exhibits significant advantages over \ce{TmMgGaO4}. 
The absence of Mg/Ga disorder in \ce{NaTmSe2} eliminates charge disorder, offering a more controllable environment for studying its magnetism.
Belonging to the rare-earth chalcogenide \ce{ARECh2}, \ce{NaTmSe2} features a large structural tolerance factor, allowing for substitution of ligand anions and cations without altering its crystal structure or symmetry.
This flexibility allows for tuning of Ising spin interactions, leading to the emergence of distinct spin ground states and exotic magnetic excitations. 

We present a comprehensive study of \ce{NaTmSe2} to uncover its fundamental magnetism.
Single crystals and polycrystalline samples of \ce{NaTmSe2}, \ce{NaLuSe2}, and \ce{NaTm_{0.1}Lu_{0.9}Se2} were successfully synthesized~\cite{SM}.
By applying the CEF theory to inelastic neutron scattering (INS) experiments, we determine the CEF parameters, energy levels, and wave functions, which are further validated by thermodynamic measurements.
A symmetry analysis of the CEF wave functions reveals that the magnetism in \ce{NaTmSe2} is well described by a $J_{1}$-$J_{2}$ TFIM, where  the transverse field term corresponds to the energy gap between the CEF ground- and first excited states.
Based on the TFIM, we investigate the ground state and spin excitations of \ce{NaTmSe2}. 
Neutron scattering and $\mu$SR measurements down to millikelvin temperatures reveal a complex ground state, where the multipolar component is spin-polarized (or paramagnetic), while the dipolar component remains disordered due to strong quantum fluctuations.
INS reveals gapless and continuous excitations mediated by the multipolar transverse field. 
Low-temperature specific heat and numerical calculations further confirm that these excitations are gapless spinons emerging from the dipolar disordered ground state.

\textit{CEF excitations of \ce{NaTmSe2}}---
\ce{Tm^{3+}} with a 4$f^{12}$ configuration has the $^{3}H_{6}$ and $^{3}H_{4}$ spectral terms, with 13- and 9-fold degeneracies, respectively, after SOC.
About 700 meV~\cite{PETROV2019103} energy gap between these terms justifies focusing on the lower $^{3}H_{6}$ term for magnetic studies below room temperature.
In the $D_{3d}$ CEF environment, the $^{3}H_{6}$ degeneracy is partially lifted. 
The symmetry of the CEF wave functions is described by the double group with the $SU(13)$ symmetry in the angular momentum $\hat{J}$-space.
Double group analysis classifies the CEF wave functions into three $\Gamma_{1}$ singlets, two $\Gamma_{2}$ singlets, and four $\Gamma_{3}$ doublets \cite{SM}.
Unlike the Kramers ion \ce{Yb^{3+}}, \ce{Tm^{3+}},which has an even number of 4$f$-electrons, lacks Kramers doublets protected by time-reversal symmetry \cite{PhysRevResearch.2.043013}.
This key distinction gives rise to the unique Ising magnetism in \ce{NaTmSe2}, as shown in Fig.~\ref{Fig:Intro}\textbf{(b)} and \textbf{(c)}. 
Further details are discussed below.

\begin{figure}[t!]
	\includegraphics[angle=0,width=1\linewidth]{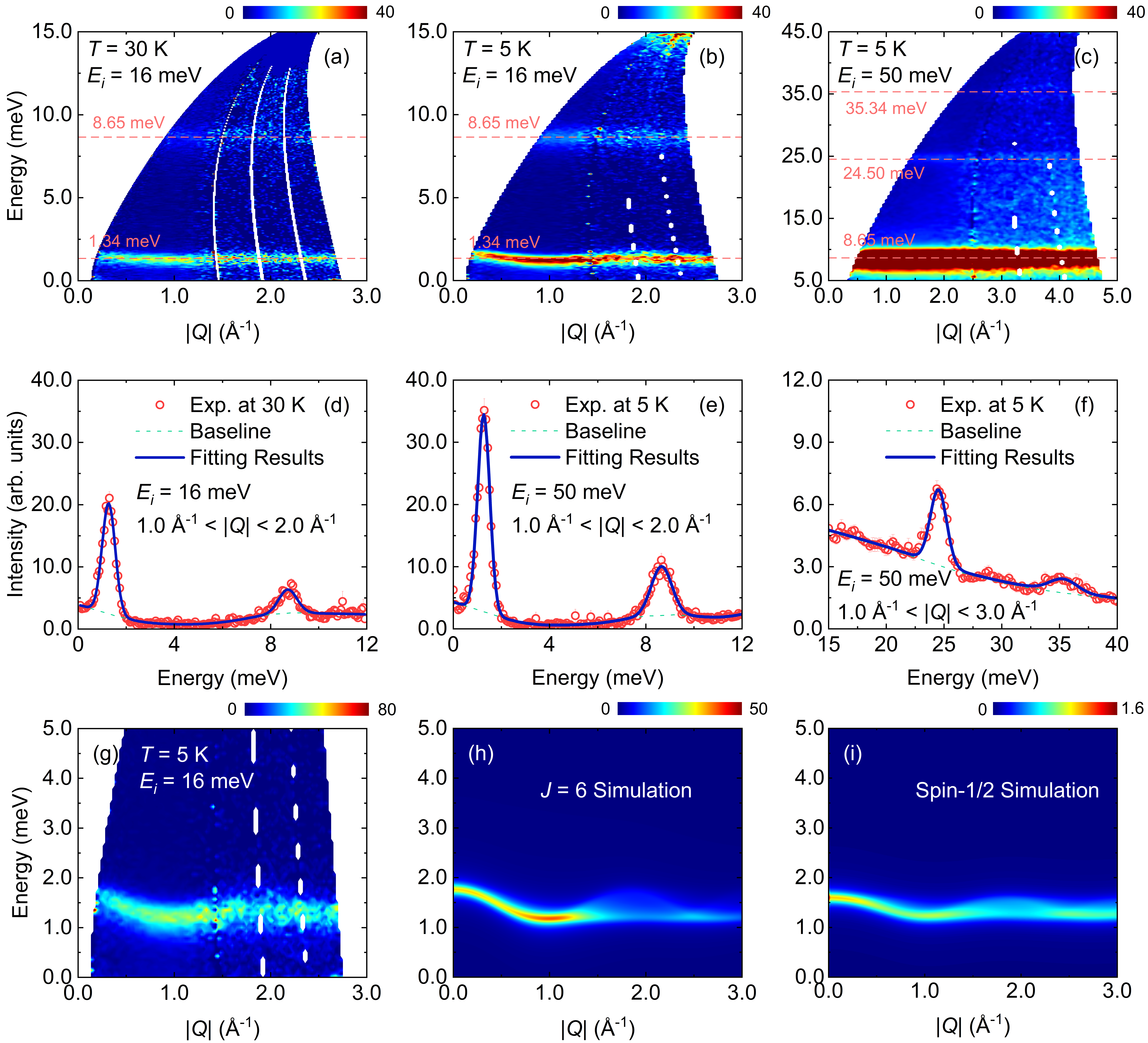}
	\renewcommand{\figurename}{\textbf{Fig. }}
	\caption{\textbf{INS spectra of polycrystalline samples of \ce{NaTmSe2}.} 
		The INS spectra measured on polycrystalline samples with incident neutron energies $E_{i}$ $=$ 16 meV at 30 K \textbf{(a)}, $E_{i}$ $=$ 16 meV at 5 K \textbf{(b)}, and $E_{i}$ $=$ 50 meV at 5 K \textbf{(c)}. Pink dashed lines mark the observed CEF excitations at 1.34, 8.65, 24.50, and 35.34 meV.
		Experimental data (red circles) in \textbf{(d)} and \textbf{(e)} were integrated over $|\vec{Q}| = \left[1.0, 2.0\right]$ \AA$^{-1}$, and in \textbf{(f)} over $|\vec{Q}| = \left[1.0, 3.0\right]$ \AA$^{-1}$.
		The fitting curves (blue lines) are based on CEF theory and the INS dipole approximation.
		\textbf{(g)} highlights the first CEF excitation with $|\vec{Q}|$-dependence.
		The $SU(13)$ spin wave simulation ($J$ = 6) and the effective spin-1/2 model successfully reproduce the excitation spectrum in \textbf{(h)} and \textbf{(i)}.
	}
	\label{Fig:NaTmSe2CEF}
\end{figure}

Understanding CEF excitations is crucial for rare-earth magnetism.
INS experiments on polycrystalline \ce{NaTmSe2} and reference \ce{NaLuSe2} \cite{SM} were performed on the MAPS spectrometer at ISIS  facility~\cite{neutron1} with incident neutron energies $E_{i}$ = 16.0 and 50.0 meV at 30 K and 5 K (Fig.~\ref{Fig:NaTmSe2CEF}\textbf{(a)}-\textbf{(c)}).
Four excitations are identified at 1.34, 8.65, 24.50, and 34.34 meV. 
The lowest one exhibits slight dispersion at 5 K, which becomes smeared out at 30 K, while the others display no $|\vec{Q}|$-dependence.
This is related to spin-spin interactions.
By applying the CEF theory and the dipole approximation to fit the INS spectra integrated over $|\vec{Q}|$ ranges $\left[1.0, 2.0\right]$ \AA$^{-1}$ and $\left[1.0, 3.0\right]$ \AA$^{-1}$ (Fig.~\ref{Fig:NaTmSe2CEF}\textbf{(d)}-\textbf{(f)}), 
we determine the CEF parameters, energy levels, and wave functions of \ce{NaTmSe2} \cite{SM}, providing key insights into its magnetism.
(a) The determined CEF wave functions are consistent with the double group analysis mentioned above.
(b) The first and second CEF excitation levels are located at 1.34 meV and 8.67 meV, respectively. 
The substantial separation that helps prevent interference from higher CEF states when constructing the low-energy spin Hamiltonian.
In comparison, \ce{KTmSe2} has its first two CEF levels at 1.17 meV and 2.82 meV \cite{RN124}.
(c) The CEF excitations in \ce{NaTmSe2} significantly differ from those in \ce{TmMgGaO4}, despite sharing the same CEF symmetry. In \ce{TmMgGaO4}, 
the first CEF excitation level at $\sim$0.5–0.6 meV \cite{RN109,RN122} merges with the spin wave excitations.

The determined CEF information is validated by thermodynamic measurements.
As shown in Fig.~\ref{Fig:Intro}\textbf{(b)} and \textbf{(c)}, the calculated CEF magnetization~\cite{SM} agrees well with the experimental data for the diluted magnetic compound \ce{NaTm_{0.1}Lu_{0.9}Se2} along both the \textit{a}- and \textit{c}-axis.
The magnetization of \ce{NaTmSe2} along the \textit{a}-axis matches that of \ce{NaTm_{0.1}Lu_{0.9}Se2}, while its suppression along the \textit{c}-axis at low temperatures indicates antiferromagnetic Ising interactions in \ce{NaTmSe2}.
The suppression of magnetization along the \textit{c}-axis is also reflected in the zero-field (ZF) specific heat, which exhibits a broad peak near 8 K (open red squares in Fig.~\ref{Fig:Intro}\textbf{(f)}).
While CEF calculations (red line) reproduce this feature, they overestimate the experimental data,
suggesting that spin interactions suppress the release of magnetic entropy. 
With applied fields along the \textit{c}-axis, this suppression weakens, and the CEF calculations align more closely with the experimental results (green and blue data, Fig.~\ref{Fig:Intro}\textbf{(e)}).

\textit{$J_{1}$-$J_{2}$ TFIM}---
The determined CEF states provide insight into the Ising magnetism of \ce{NaTmSe2}, as revealed by thermodynamic measurements. 
The magnetic Hamiltonian is typically expressed in terms of the total angular momentum $J = 6$ as follows.
\begin{equation}
	\hat{H}_{eff} = \hat{H}_{CEF} + \varTheta_{1} \sum_{\left\langle i,j \right\rangle} \hat{J}^{z}_{i} \hat{J}^{z}_{j} + \varTheta_{2} \sum_{\left\langle \left\langle i,k \right\rangle \right\rangle} \hat{J}^{z}_{i} \hat{J}^{z}_{k}
\label{Eq:TotalHamiltonian}
\end{equation}
Here, $\hat{H}_{CEF}$ is the CEF Hamiltonian \cite{SM}, and $\hat{J}_{i}^{z}$ represents the Ising spin with angular momentum $J = 6$.
The nearest neighbor (NN) and NNN Ising interactions are denoted as $\varTheta_{1}$ and $\varTheta_{2}$, respectively. 
This Hamiltonian captures the slight $|\vec{Q}|$-dependence of the lowest CEF excitation arising from Ising interactions. 
Simulations yield optimal parameters $\varTheta_{1} = 0.00354(2)$ meV and $\varTheta_{2} = 0.00046(1)$ meV, successfully reproducing the INS spectrum (Fig.~\ref{Fig:NaTmSe2CEF}(\textbf{h})).

Moreover, the second CEF level ($\sim$8.7 meV) lies significantly above the first ($\sim$1.3 meV), enabling the Hamiltonian to be reduced to a low-energy effective model involving only the lowest two CEF states to describe the spin physics at low temperatures.
The experimentally determined wave functions for the CEF ground state $\left| {{\psi _0}} \right\rangle$ and the first excited state $\left| {{\psi _1}} \right\rangle$ are given as follows:
\begin{equation}
	\label{Eq:CEFGand1Exp}
	\begin{array}{l}
		\left| {{\psi _0}} \right\rangle  =  - 0.52\left( {\left| 6 \right\rangle  + \left| { - 6} \right\rangle } \right) + 0.38\left( { - \left| 3 \right\rangle  + \left| { - 3} \right\rangle } \right) + 0.42\left| 0 \right\rangle \\
		\\
		\left| {{\psi _1}} \right\rangle  = 0.58\left( {\left| 6 \right\rangle  - \left| { - 6} \right\rangle } \right) + 0.40\left( {\left| 3 \right\rangle  + \left| { - 3} \right\rangle } \right)
	\end{array}
\end{equation}
The wave functions form a non-Kramers doublet, which can be represented as effective spin-1/2 ``up'' and ``down'' states through recombination.
This symmetry reduces the complex CEF system to a simplified effective spin-1/2 model, capturing its magnetic behavior and, in particular, the Ising-type interactions \cite{SM}.

Therefore, the low-energy effective spin-1/2 $J_{1}$-$J_{2}$ TFIM for \ce{NaTmSe2} is given by \cite{PhysRevX.10.011007,RN109}.
\begin{equation}
	\hat{H} = J_{1} \sum_{\left\langle ij \right\rangle} S_{i}^{z} S_{j}^{z} + J_{2} \sum_{\left\langle \left\langle  ik \right\rangle \right\rangle} S_{i}^{z} S_{k}^{z} -\Delta \sum_{i} S_{i}^{y}
\label{Eq:OneHalfHamiltonian}
\end{equation}
Here, $J_{1}$ and $J_{2}$ represent the Ising interactions between dipolar $S_{i}^{z}$,
while the energy gap $\Delta = 1.34$ meV between the CEF ground and first excited states maps onto a transverse field acting on the multipolar $S_{i}^{y}$.

At low-energies, the TFIM is equivalent to the spin Hamiltonian described by Eq.~\ref{Eq:TotalHamiltonian}.
The parameters $\varTheta_{1}$ and $\varTheta_{2}$ convert to $J_{1}$ and $J_{2}$ via a factor $\frac{J(J+1)}{S(S+1)} = 56$ \cite{PhysRevB.101.144432,SM}, yielding $J_{1} = \varTheta_{1} \times 56 \simeq 0.198$ meV ($\sim$ 2.3 K) and $J_{2} = \varTheta_{2} \times 56 \simeq 0.026$ meV ($\sim$ 0.3 K). 
Using the TFIM, we successfully reproduce the INS spectrum near 1.34 meV, as shown in Fig.~\ref{Fig:NaTmSe2CEF}\textbf{(i)}, confirming the equivalence of the two models at low energies.

Furthermore, the magnetic moment along the $S^{z}$ direction is given by $-2 \cdot g_J \mu_B S^{z} \left\langle \psi_0 \middle| \hat{J_z} \middle| \psi_1 \right\rangle$~\cite{PhysRevX.10.031069} ($g_J = \frac{7}{6}$ for \ce{Tm^{3+}}).
This yields an effective $g$-factor $g_{\text{eff}} = 2 \cdot g_J \left\langle \psi_0 \middle| \hat{J_z} \middle| \psi_1 \right\rangle \sim$10.55 along the \textit{c}-axis, consistent with the magnetization data in Fig.~\ref{Fig:Intro}\textbf{(e)}.
In contrast, the in-plane $g$-factor $g_J \left\langle \psi_0 \middle| \hat{J_\alpha} \middle| \psi_1 \right\rangle$ ($\alpha$ = $x$ or $y$) vanishes. This is confirmed by the absence of a saturation moment in Fig.~\ref{Fig:Intro}\textbf{(e)}, where only a linear paramagnetic contribution from CEF excitations exists.

The distinct ground states of \ce{NaTmSe2} and \ce{TmMgGaO4} can be understood through their TFIM parameters.
In \ce{TmMgGaO4}, $J_{1}$ ($\sim$0.54 meV) is comparable to the transverse field ($\sim$0.63 meV) \cite{RN109}, leading to a resonance-like coupling and a magnetically ordered ground state.
In \ce{NaTmSe2}, the transverse field is larger than $J_{1}$, causing multipolar moments to freeze at low temperatures~\cite{PhysRevB.68.104409}.
Considering the noncommutativity of $S^{z}$ and $S^{y}$ in TFIM, this leads to an unusual spin ground state where the dipolar component remains spin-disordered due to Ising interactions, while the multipolar component is paramagnetic (or polarized).

\begin{figure}[t!]
	\includegraphics[angle=0,width=1\linewidth]{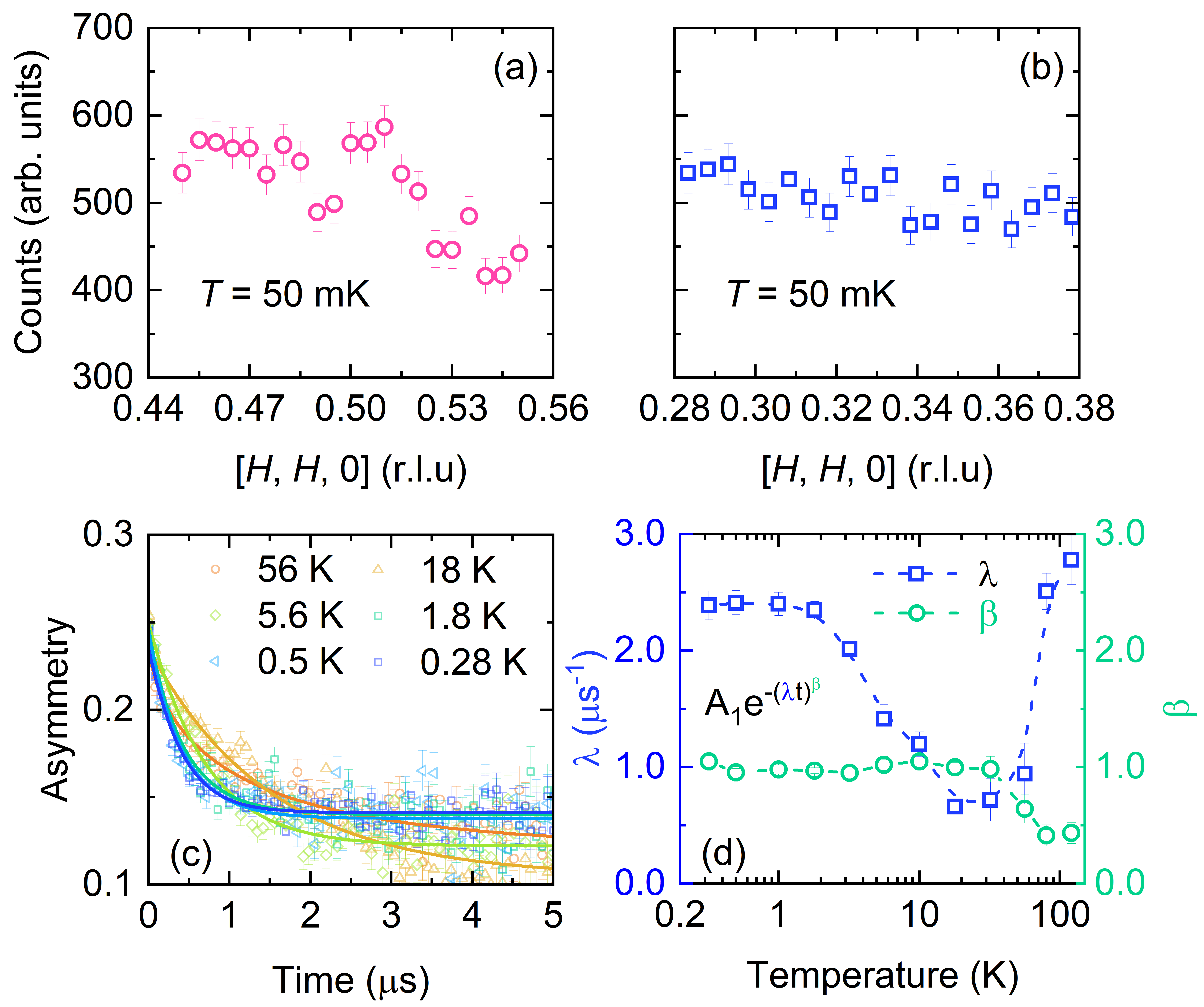}
	\renewcommand{\figurename}{\textbf{Fig. }}
	\caption{\textbf{Neutron diffraction and ZF-$\mu$SR experiments.} 
		\textbf{(a)}--\textbf{(b)} Neutron diffraction results for single-crystal \ce{NaTmSe2} around the wave vectors $\vec{Q} = [0.5, 0.5, 0]$ and $[1/3, 1/3, 0]$ at 50 mK.  
		\textbf{(c)} ZF-$\mu$SR asymmetry spectra at various temperatures.  
		\textbf{(d)} Temperature dependence of the ZF-$\mu$SR spin relaxation rate $\lambda$ (blue squares) and stretching exponent $\beta$.
	}
	\label{Fig:DiffAndMusr}
\end{figure}
\textit{Spin ground state}---
To investigate the ground state, we performed neutron diffraction~\cite{neutron2} on a single crystal of \ce{NaTmSe2} at the high-symmetry points $M$ $\left( \left[1/2, 1/2, 0\right] \right)$ and $K$ $\left( \left[1/3, 1/3, 0\right] \right)$.
No magnetic diffraction peaks were observed at 50 mK (Fig.~\ref{Fig:DiffAndMusr}\textbf{(a)}-\textbf{(b)}) or 0.85 K \cite{SM}.

We further used $\mu$SR~\cite{msr} to investigate the ground state of \ce{NaTmSe2}. 
The ZF-$\mu$SR asymmetry spectra (Fig.~\ref{Fig:DiffAndMusr}\textbf{(c)}) are described by a model of stretched exponential (SE) decay.
\begin{equation}
	A\left(t\right) = A_{0} + A_{1}  e^{-\left(\lambda t \right)^{\beta}}
\end{equation}
where $A_{0}$ is a constant asymmetry fraction within time window, $A_{1}$ is the initial asymmetry for the SE component, $\lambda$ is the $\mu^{+}$ spin relaxation rate, and $\beta$ is the stretching exponent.
No oscillations are observed down to 0.28 K, similar to other QSL candidates, such as \ce{YbMgGaO4} \cite{PhysRevLett.117.097201}, \ce{NaYbO2} \cite{PhysRevB.100.144432}, \ce{NaYbS2} \cite{PhysRevB.100.241116}, and \ce{NaYbSe2} \cite{PhysRevB.106.085115}.
This excludes a magnetically ordered ground state, consistent with the neutron diffraction results. 
We also notice the growth of the relaxation tail with decreasing temperature in Fig.~\ref{Fig:DiffAndMusr} (c).
The long-time tail is commonly observed in materials with spin freezing, such as the typical spin glass material \ce{NiGa2S4}~\cite{PhysRevB.78.220403}. However, in \ce{NaTmSe2}, this feature is associated with quantum fluctuations induced by the transverse field at low temperatures.
For example, similar long-time tail features are observed in \ce{PrZnAl11O9}~\cite{PhysRevB.109.165143} and \ce{Pr3BWO9}~\cite{PhysRevResearch.6.023267}, which also exhibit TFIM magnetism.

The ZF $\mu^{+}$ spin relaxation rate $\lambda$, which probes local spin fluctuations, exhibits a non-monotonic temperature dependence. 
$\lambda$ is enhanced at higher temperatures due to thermal effects and decreases continuously upon cooling.
Below 3 K, however, it saturates into a plateau.
Since muons probe low-energy spin fluctuations through dipole interactions, this indicates that quantum fluctuations arising from dipolar Ising interactions drive the spin system into a disordered state.
The stretch exponent $\beta$ remains around 1.0 between 0.3 and 38 K (Fig.~\ref{Fig:DiffAndMusr}\textbf{(d)}), 
indicating that the $\mu$SR spectra of \ce{NaTmSe2} are well described by a simple exponential decay within this temperature range.
This suggests that the muon probes a homogeneous, dynamically evolving local magnetic environment, similar to the dynamic, gapless excitations observed in \ce{YbMgGaO4} \cite{PhysRevLett.117.097201}.

\begin{figure}[t!]
	\includegraphics[angle=0,width=1\linewidth]{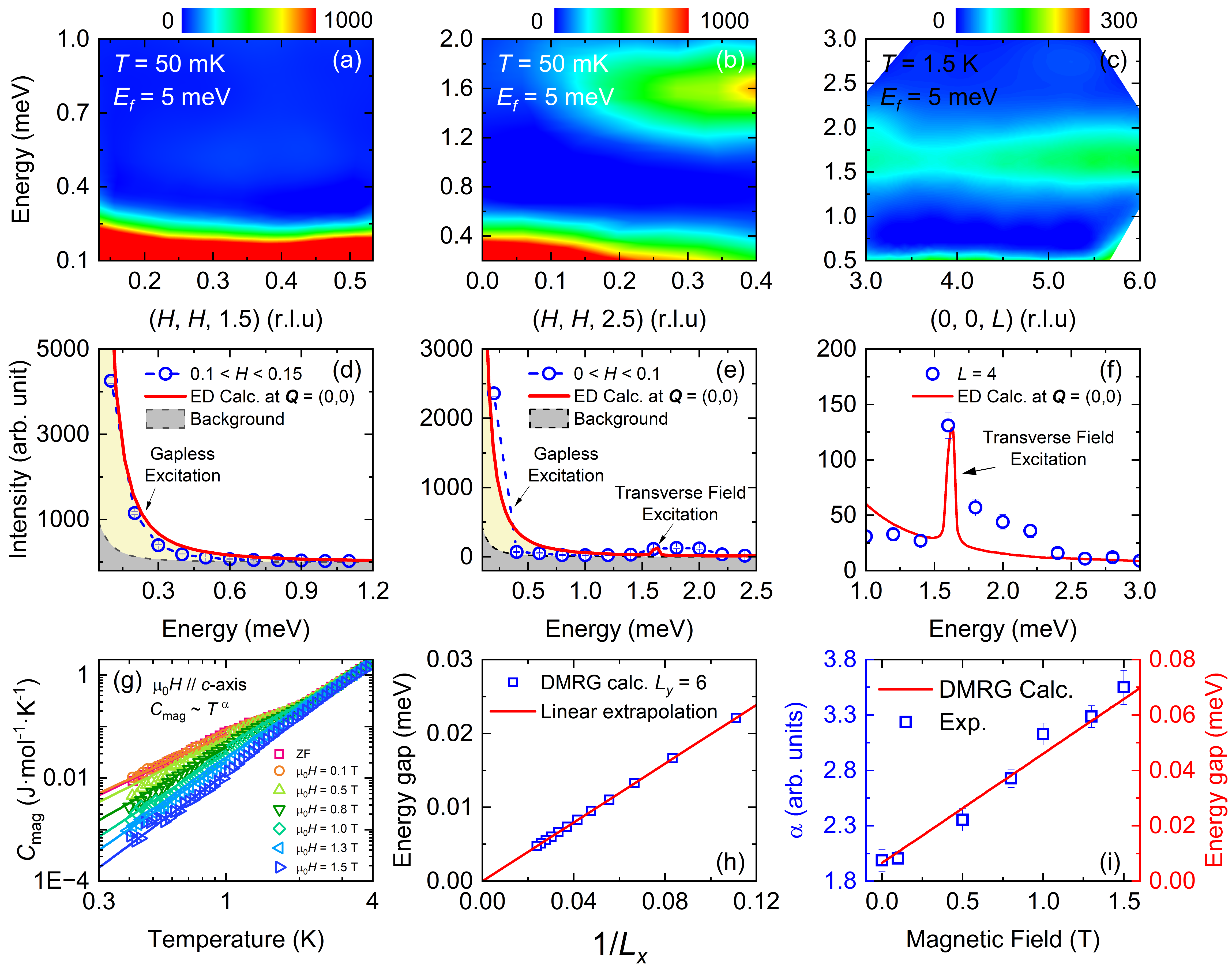}
	\renewcommand{\figurename}{\textbf{Fig. }}
	\caption{\textbf{Spin excitations of \ce{NaTmSe2}.} 
		\textbf{(a)}–\textbf{(b)} INS spectra measured at 50 mK with fixed final neutron energy $E_{f}$ $=$ 5 meV along $[H, H, 1.5]$ and $[H, H, 2.5]$, respectively. The excitation near 1.6 meV in (b) corresponds to the transverse field term from the first CEF excitation.
		\textbf{(c)} INS spectrum measured at 1.5 K with fixed final neutron energy $E_{f}$ $=$ 5 meV along $[0, 0, L]$.
		\textbf{(d)}-\textbf{(e)} Energy dependence of INS intensity near $\vec{Q} = [0, 0, 1.5]$ and $\vec{Q} = [0, 0, 2.5]$ (blue circles), respectively, compared with ED calculations of the dynamic structure factor at  $\vec{Q} = [0, 0]$ (red line). The instrumental resolution (black dashed line) defines the elastic background.
		\textbf{(f)} Energy dependence of INS intensity at $\vec{Q} = [0, 0, 4]$ (blue circles), compared with ED calculation.
		\textbf{(g)} Magnetic specific heat of \ce{NaTmSe2} under different magnetic fields along the \textit{c}-axis at low temperatures, showing power-law behavior $C_{mag} \sim T_{\alpha}$ (solid lines through data points).
		\textbf{(h)} DMRG calculations of the ground state energy gap versus system size $L_{x}$ (at fixed $L_{y}$ $=$ 6).  
		\textbf{(i)} Magnetic field dependence of the specific heat power-law exponent \(\alpha\) (blue squares, experimental) and energy gap (red line, DMRG calculations for 6 $\times$ 30 system).
	}
	\label{Fig:Spinon}
\end{figure}
\textit{Spin excitations}---
Elastic neutron scattering and $\mu$SR experiments rule out magnetic ordering in the ground state of \ce{NaTmSe2}, while the spin relaxation rate $\lambda$ indicates the presence of strong spin fluctuations. 
To probe low-energy spin excitations, we conducted INS~\cite{neutron2} experiments along $\left[H, H, 1.5\right]$ (Fig.~\ref{Fig:Spinon}\textbf{(a)}) and $\left[H, H, 2.5\right]$ (Fig.~\ref{Fig:Spinon}\textbf{(b)}) at 50 mK, resolving continuous excitations in the 0.1 - 0.5 meV range \cite{SM}. 
The spectral cutoff near $\vec{Q} = \left[0, 0, 1.5\right]$ and $\left[0, 0, 2.5\right]$ ( Fig.~\ref{Fig:Spinon}\textbf{(d)} and \textbf{(e)} ) clearly reveals distinct gapless spin excitations, which rise well above the elastic background, as shown by the black dashed line~\cite{SM}. Considering the spectrometer’s energy resolution of 0.13 meV~\cite{SM} and the measured elastic background intensity of $I$ = 1503.74~\cite{SM}, the observed excitations are well-resolved and cannot be attributed to residual elastic scattering. These factors further reinforce the intrinsic nature of the gapless spin excitations in \ce{NaTmSe2}.

Exact diagonalization (ED) and dynamic Green's function calculations \cite{KAWAMURA2017180,IDO2024109093} offer further numerical insight into the gapless low-energy spin excitations \cite{SM}.
We calculated the spin excitation spectrum at $\Gamma$ point for a triangular spin lattice ($L_{x} \times L_{y}$ = 6 $\times$ 4) with periodic boundary conditions, as shown by the red line in Fig.~\ref{Fig:Spinon}\textbf{(d)} and \textbf{(e)}.
The results reveal dominant low-energy excitations below 0.5 meV, underscoring the significant contribution of gapless modes to spin dynamics. 
Furthermore, the excitation around 1.6 meV attributed to the multipolar transverse field, as obtained from ED calculations, is consistent with the INS results measured along the $\left[0, 0, L\right]$ direction (see Fig.~\ref{Fig:Spinon}\textbf{(c)} and \textbf{(f))}.
Overall, our ED calculations show agreement with the INS data.

A key question concerns the origin of the low-energy gapless excitations, which are attributed to spinon excitations mediated by the multipolar transverse field.
Similar to a general TFIM, the noncommutativity between $S^{z}$ and $S^{y}$, constructed from the CEF ground and first excited states, induces strong quantum fluctuations that bridge the flipping between the ``up" and ``down" states of the dipolar moment, leading to a spin-disordered (or QSL) state.
Notably, $S^{y}$ here acts as a multipolar operator rather than a dipolar one, offering new insights into the control of quantum fluctuations.
Besides, a subtle enhancement of magnetic diffuse scattering at $\vec{Q}$ = $\left(1/3, 1/3, L\right)$ ($L$ = $\left(0, 1, 2\right)$) upon cooling~\cite{SM}, as compared to $\left(1/2, 1/2, L\right)$, is consist with short-range antiferromagnetic correlations expected from triangular lattice Ising system under strong transverse fields.

Low-temperature specific heat measurements further support this interpretation. 
As shown in Fig.~\ref{Fig:Spinon}\textbf{(g)}, the ZF specific heat below 1.5 K follows a power-law behavior, $C_{p}\left(T\right) \sim T^{\alpha}$, with a fitted exponent $\alpha$ = 1.99 $\pm$ 0.10 (RMSE = 0.0043). 
Under a 0.1 T field applied along the \textit{c}-axis, $\alpha$ is 2.01 $\pm$ 0.06 (RMSE = 0.0026).
The proximity of $\alpha$ to 2 is nontrivial, suggesting a gapless QSL state where low-energy spin excitations are spinons \cite{PhysRevLett.98.117205, PhysRevLett.123.207203}.
Similar gapless excitations have been reported in the Pr-based triangular Ising magnet \ce{PrMgAl11O19} \cite{PhysRevB.109.165143}. Increasing the magnetic field along the c-axis causes the low-temperature specific heat to deviate from the $\sim$ $T^{2}$
power law, suggesting that the field suppresses spinon excitations driven by quantum fluctuations, thereby driving the system from a gapless to a gapped state.

To gain deeper insight into the low-energy excitations, we employ density matrix renormalization group (DMRG) calculations.
As shown in Fig.~\ref{Fig:Spinon}\textbf{(h)}, for $L_{y} = 6$, we compute the energy gap between the ground and first excited states as a function of $L_{x}$ \cite{SM}.
The results reveal that the energy gap decreases as $L_{x}$ ($1/L_{x}$) increases (decreases).
From a linear extrapolation, we infer that the spin system is gapless in the thermodynamic limit.
By calculating the field dependence of the energy gap for a spin lattice of $L_{y} \times L_{x} = 6 \times 30$ (red line, Fig.~\ref{Fig:Spinon}\textbf{(i)}), we find that the gap increases linearly with the field, consistent with the linear behavior of the power-law exponent $\alpha$ (open blue squares).
Cross-analysis of INS spectra, low-temperature specific heat, and numerical calculations confirms that the low-energy excitations in \ce{NaTmSe2} are gapless spinons driven by the transverse field.

\textit{Summary}---
We investigate the unusual Ising magnetism, CEF excitations, ground state, and spin dynamics of \ce{NaTmSe2}.
INS measurements reveal four CEF excitation levels. 
By fitting the spectra using CEF theory, we determine the CEF parameters, energy levels, and wave functions, which are validated through simulations of the magnetization data for the dilute reference crystal \ce{NaTm_{0.1}Lu_{0.9}Se2}.
Using the general magnetic Hamiltonian with $J = 6$ and the NN and NNN Ising spin interactions, 
we accurately reproduce the slight dispersion of the first CEF excitation and determine the spin interaction parameters.
Based on the CEF ground and first excited states, we construct an effective spin-1/2 $J_{1}$-$J_{2}$ TFIM, where the dipole $S^{z}$ operators exhibit Ising interactions, and the transverse field acts on the multipolar $S^{y}$.
We experimentally investigate the ground state and low-energy spin excitations. 
Elastic neutron scattering reveals no magnetic Bragg peaks down to 50 mK, while ZF-$\mu$SR confirms a spin-disordered ground state with strong fluctuations. 
The low-energy INS continuum below 0.5 meV and the near power-law behavior of the ZF specific heat suggest gapless spinon excitations, originating from the dipolar spin-disordered ground state and mediated by the multipolar transverse field.

\textit{Acknowledgements}---
We gratefully acknowledge Rong Yu , Zhengxin Liu , Zhentao Wang, and Long Zhang for helpful discussion.
This work was supported by 
the National Key Research and Development Program of China (Grant Nos. 2024YFA1408300 and 2022YFA1402704), 
the National Science Foundation of China (Grant No. 12274186), 
the Strategic Priority Research Program of the Chinese Academy of Sciences (Grant No. XDB33010100), 
and the Synergetic Extreme Condition User Facility (SECUF, \href{https://cstr.cn/31123.02.SECUF}{https://cstr.cn/31123.02.SECUF} ).
We thank ISIS facility for beam time on MAPS RB1920091.
The $\mu$SR measurements were performed at the Dolly spectrometer of the
Swiss Muon Source (Paul Scherrer Institut, Villigen, Switzerland).


%

\end{document}